\documentclass[runningheads,fleqn]{svmult}
\usepackage{makeidx}
\usepackage{graphicx}
\usepackage{subeqnar}
\usepackage{multicol}
\usepackage{taphys}

\begin{document}

\title*{Scale Invariance and Dynamic Phase Transitions in Diffusion-Limited 
        Reactions}
\toctitle{Scale Invariance and Dynamic Phase Transitions in Diffusion-Limited 
          Reactions}
\titlerunning{Dynamic Phase Transitions in Diffusion-Limited Reactions}

\author{Uwe C. T\"auber}
\authorrunning{Uwe C. T\"auber}

\institute{Department of Physics, Virginia Tech, Blacksburg, VA 24061-0435, USA
           \\ email: tauber@vt.edu}
\maketitle

\begin{abstract}
Many systems that can be described in terms of diffusion-limited `chemical' 
reactions display non-equilibrium continuous transitions separating active from
inactive, absorbing states, where stochastic fluctuations cease entirely. 
Their critical properties can be analyzed via a path-integral representation 
of the corresponding classical master equation, and the dynamical 
renormalization group. 
An overview over the ensuing universality classes in single-species processes 
is given, and generalizations to reactions with multiple particle species are 
discussed as well.
The generic case is represented by the processes $A \rightleftharpoons A + A$, 
and $A \to \emptyset$, which map onto Reggeon field theory with the critical 
exponents of directed percolation (DP). 
For branching and annihilating random walks (BARW) $A \to (m+1) A$ and 
$A + A \to \emptyset$, the mean-field rate equation predicts an active state 
only. 
Yet BARW with odd $m$ display a DP transition for $d \leq 2$. 
For even offspring number $m$, the particle number parity is conserved locally.
Below $d_c' \approx 4/3$, this leads to the emergence of an inactive phase that
is characterized by the power laws of the pair annihilation process. 
The critical exponents at the transition are those of the `parity-conserving' 
(PC) universality class. 
For local processes without memory, competing pair or triplet annihilation and 
fission reactions $k A \to (k - l) A$, $k A \to (k+m)A$ with $k=2,3$ appear to
yield the only other universality classes not described by mean-field theory. 
In these reactions, site occupation number restrictions play a crucial role. \\

\end{abstract}

\section{Introduction: Active to absorbing state transitions}
\label{sec:intr}

The characterization of non-equilibrium steady states\index{non-equilibrium 
steady states} constitutes one of the prevalent goals in present statistical 
mechanics.
Unfortunately, away from thermal equilibrium one cannot in general derive even 
stationary macroscopic properties from an effective free energy function.
One might hope, however, that such a classification in terms of symmetries and 
interactions becomes feasible near continuous phase transitions separating 
different non-equilibrium steady states: 
Drawing on the analogy with equilibrium critical points, one would expect 
certain features of non-equilibrium phase transitions\index{non-equilibrium 
phase transitions} to be {\em universal} as well, i.e., independent of the 
detailed microscopic dynamical rules and even the initial conditions.
The emerging power laws and scaling functions describing the long-wavelength, 
long-time limit should then hopefully be characterized by not too many distinct
dynamic universality classes\index{universality classes}.
Yet studies of a variety of non-equilibrium processes have taught us that 
critical phenomena as well as generic scale invariance far from thermal 
equilibrium, where restrictive detailed-balance constraints do not apply, are 
often considerably richer than their equilibrium counterparts.
Indeed, intuitions inferred from the latter have frequently turned out to be 
quite deceptive.

But as in the investigation of equilibrium critical phenomena, field theory 
representations supplemented with the dynamical version of the renormalization 
group (RG)\index{renormalization group} provide powerful methods to extract and
systematically classify universal properties at continuous non-equilibrium 
phase transitions.
Additional indispensable quantitative tools are of course Monte Carlo 
simulations, and other numerical approaches and exact solutions when available 
(the latter are usually restricted to specific one-dimensional systems).
Analytical and numerical methods usually supplement each other:
Simulation results often call for deeper understanding of the underlying 
processes, but also rely on some theoretical background as a basis for data 
analysis.

A prominent class of genuine non-equilibrium phase transitions separates 
{\em `active'}\index{active states} from {\em `inactive, absorbing'} stationary
states\index{absorbing states} where, owing to the absence of any agents, 
stochastic fluctuations cease entirely \cite{chopard98,hinrichsen00}.
These occur in a variety of systems in nature; e.g., in chemical 
reactions\index{chemical reactions} which involve an inert state $\emptyset$ 
that does not release the reactants $A$ anymore.
Another example are models for stochastic population dynamics\index{population 
dynamics}, combining, say, diffusive migration of a species $A$ with asexual 
reproduction $A \to 2 A$ (with rate $\sigma$), spontaneous death 
$A \to \emptyset$ (at rate $\mu$), and lethal competition $2 A \to A$ (with 
rate $\lambda$). 
In the inactive state, where no population members $A$ are left, clearly all 
processes terminate.
Similar effective dynamics may be used to model certain non-equilibrium 
physical systems.
Consider for instance the domain wall\index{domain wall} kinetics in Ising 
chains with competing Glauber and Kawasaki dynamics \cite{grassberger84}. 
Here, spin flips $\uparrow \uparrow \downarrow \downarrow \, \to \,
\uparrow \uparrow \uparrow \downarrow$ and $\uparrow \uparrow 
\downarrow \uparrow \, \to \, \uparrow \uparrow \uparrow \uparrow$ 
may be viewed as domain wall ($A$) hopping and pair annihilation 
$2 A \to \emptyset$, whereas spin exchange $\uparrow \uparrow \downarrow 
\downarrow \, \to \, \uparrow \downarrow \uparrow \downarrow$ represents a 
branching process $A \to 3 A$. 
Notice that the para- and ferromagnetic phases respectively map onto the active
and inactive `particle' states, the latter rendered absorbing if the spin flip 
rates are computed at zero temperature, thus forbidding any energy increase.

The simplest mathematical description for such processes uses kinetic rate 
equations\index{rate equations}, which govern the time evolution of the mean 
`particle' density $n(t)$.
For example, the above population model leads to Fisher's rate equation
\begin{equation}
  \partial_t \, n(t) = \left( \sigma - \mu \right) n(t) - \lambda \, n(t)^2 \ .
\label{mfreq}
\end{equation}
It yields both inactive and active phases:
For $\sigma < \mu$ we have $n(t \to \infty) \to 0$, whereas for $\sigma > \mu$ 
the density eventually saturates at $n_s = (\sigma - \mu) / \lambda$.
The explicit solution (with initial particle density $n_0$)
\begin{equation}
  n(t) = n_0 \, n_s \Big/ \Bigl[ n_0 + (n_s - n_0) \, e^{(\mu - \sigma) t} 
  \Bigr]
\label{mfsol}
\end{equation}
shows that both stationary states are approached exponentially in time.
The two phases are separated by a continuous non-equilibrium phase transition 
at $\sigma = \mu$, where the temporal decay becomes algebraic, 
$n(t) = n_0 / (1 + n_0 \lambda \, t) \to 1/(\lambda t)$ as $t \to \infty$, 
independent of the initial density.
But Eq.~(\ref{mfreq}) represents a mean-field approximation\index{mean-field 
approximation}, as we have in fact replaced the joint probability of finding 
two particles at the same position with the square of the mean density.
As in equilibrium, however, critical fluctuations\index{critical fluctuations} 
are expected to invalidate simple mean-field theory in sufficiently low 
dimensions $d < d_c$, which defines the upper critical dimension\index{upper 
critical dimension}.
A more satisfactory treatment therefore necessitates a systematic incorporation
of spatio-temporal fluctuations, including specifically the particle 
correlations as induced by the dynamics.

\section{From the master equation to stochastic field theory}
\label{sec:meft}

The renormalization group study of (near-){\em equilibrium} dynamical critical 
phenomena relies on phenomenological Langevin equations\index{Langevin 
equation} for the order parameter and `slow' hydrodynamic variables associated 
with conservation laws \cite{hohenberg77}.
All other degrees of freedom are treated as Gaussian white noise, whose second 
moment is related to the relaxation coefficients through Einstein's 
fluctuation-dissipation theorem.
As we shall see, however, such a description is not necessarily possible at all
in reaction-diffusion systems.
To the very least one would have to invoke fundamental conjectures on the 
noise correlators; but far from equilibrium these often crucially influence 
long-wavelength properties.
It is therefore desirable to construct a long-wavelength effective theory for 
stochastic processes directly from their microscopic definition, without 
recourse to any serious additional assumptions or approximations.

Fortunately, there exists an established procedure to derive the Liouville time
evolution operator\index{Liouville operator} for locally interacting particle
systems immediately from the classical master equation\index{master equation}, 
wherefrom a field theory\index{field theory} representation is readily
obtained \cite{doi76}.
The key point is that all possible configurations can be labeled by specifying 
the occupation numbers $n_i$ of, say, the sites of a $d$-dimensional lattice.
Let us for now assume that there are no site occupation restrictions, i.e., any
$n_i = 0, 1, 2, \ldots$ is allowed (we shall address effects of particle 
exclusions in Sec.~\ref{sec:anfr}).
The master equation governs the time evolution of the configurational 
probability $P(\{ n_i \};t)$.
For example, the corresponding contribution from the binary coagulation process
$2 A \to A$ at site $i$ reads 
\begin{equation}
  \partial_t P(n_i;t) |_\lambda = 
  \lambda \Bigl[ (n_i+1) n_i P(n_i+1;t) - n_i (n_i-1) P(n_i;t) \Bigr] \ . 
\label{maseq}
\end{equation}
This sole dependence on the integer variables $\{ n_i \}$ calls for a 
representation in terms of bosonic ladder operators with the standard 
commutation relations $[ a_i , a_j^\dagger ] = \delta_{ij}$, and the empty 
state $| 0 \rangle$ such that $a_i | 0 \rangle  = 0$.
We next define the Fock states\index{Fock states} via 
$| \{ n_i \} \rangle = \prod_i (a_i^\dagger)^{n_i} | 0 \rangle$ (notice the 
different normalization from standard quantum mechanics), and thence construct 
the formal {\em state vector} 
\begin{equation}
  | \Phi(t) \rangle = \sum_{\{ n_i \}} P(\{ n_i \};t) \ | \{ n_i \}
  \rangle \ .
\label{stvec}
\end{equation}
The linear time evolution imposed by the master equation can be cast into an 
`imaginary-time Schr\"odinger' equation 
$\partial_t | \Phi(t) \rangle = - H \, | \Phi(t) \rangle$ with a generally 
non-Hermitian, local `Hamiltonian' $H(\{a_i^\dagger\},\{a_i\})$.
For instance, the on-site coagulation reaction is encoded in this formalism via
$H^\lambda_i = - \lambda \, (1 - a_i^\dagger) \, a_i^\dagger a_i^2$.

Our goal really is to evaluate time-dependent statistical averages for 
observables $F$, necessarily also just functions of the occupation numbers, 
$\langle F(t) \rangle = \sum_{\{ n_i \}} F(\{ n_i \}) \, P(\{ n_i \};t)$.
Straightforward algebra using the identity $[e^a,a^\dagger] = e^a$
shows that this average can be written as a `matrix element' 
\begin{equation}
  \langle F(t) \rangle = \langle {\cal P}| \, F(\{ a_i \}) \, |\Phi(t) \rangle 
  = \langle {\cal P}| \, F(\{ a_i \}) \, e^{- H t} \, |\Phi(0) \rangle
\label{avmel}
\end{equation}
with the state vector $| \Phi(t) \rangle$ and the projector state 
$\langle {\cal P} | = \langle 0 | \prod_i e^{a_i}$; notice that
$\langle {\cal P} | 0 \rangle = 1$.
Probability conservation implies 
$1 = \langle {\cal P} | \, e^{- H t} \, | \Phi(0) \rangle$, i.e., for 
infinitesimal times $\langle {\cal P} | \Phi(0) \rangle = 1$ and 
$\langle {\cal P} | H = 0$, which is satisfied provided 
$H(\{ a_i^\dagger \to 1 \} , \{ a_i \}) = 0$.
We remark that commuting the factors $e^{a_i}$ to the right has the effect of 
shifting all $a_i^\dagger \to 1 + a_i^\dagger$.
One may then emply coherent states\index{coherent states}, as familiar from 
quantum many-particle theory \cite{popov81}, to represent the matrix element 
(\ref{avmel}) as a {\em functional integral} with statistical weight 
$\exp (- S)$.
Omitting contributions from the initial state, the action becomes
\begin{equation}
  S[\{ {\hat \psi}_i \},\{ \psi_i \}] = \int \! dt \, \Bigl[ 
  \sum_i {\hat \psi}_i \, \partial_t \psi_i + H(\{{\hat \psi}_i\},\{\psi_i\}) 
  \Bigr] \ .
\label{ftact}
\end{equation}
Taking the continuum limit finally yields the desired field theory that 
faithfully encodes all stochastic fluctuations.

\section{Reggeon field theory and directed percolation (DP)}
\label{sec:dper}

Let us now return to our population dynamics example with random walkers $A$ 
(with diffusion constant $D$ in the continuum limit), subject to the Gribov 
reactions $A \rightleftharpoons A + A$ and $A \to \emptyset$.
The corresponding field theory (\ref{ftact}) reads
\begin{equation}
  S = \! \int \! d^dx \, dt \Bigl[ {\hat \psi} (\partial_t - D \nabla^2) \psi 
  + \sigma (1 - {\hat \psi}) {\hat \psi} \psi - \mu (1 - {\hat \psi}) \psi 
  - \lambda (1 - {\hat \psi}) {\hat \psi} \psi^2 \Bigr] \, . \ 
\label{dpmft}
\end{equation}
The stationarity condition $\delta S / \delta \psi = 0$ is always solved by
$\hat \psi = 1$; upon inserting this into $\delta S / \delta {\hat \psi} = 0$, 
and identifying $n(t) = \langle \psi(\vec{x},t) \rangle$, one recovers 
Fisher's mean-field rate equation (\ref{mfreq}).
Higher moments of the field $\psi$, however, cannot be immediately connected 
with density correlation functions.
In terms of an arbitrary momentum scale $\kappa$, we record the naive scaling 
dimensions $[{\hat \psi}] = \kappa^0$, $[\psi] = \kappa^d$, 
$[\sigma] = \kappa^2 = [\mu]$, and $[\lambda] = \kappa^{2-d}$.
Hence the decay and branching rates constitute relevant operators in the RG 
sense, whereas the annihilation process becomes marginal at $d = 2$.
Next we expand the action (\ref{dpmft}) about the stationary 
solution $\hat \psi = 1$, i.e., introduce 
${\tilde \psi}(\vec{x},t) = {\hat \psi}(\vec{x},t) - 1$, whereupon we arrive at
\begin{equation}
  S = \! \int \! d^dx \, dt \Bigl[ {\tilde \psi} (\partial_t - D \nabla^2) \psi
  + (\mu - \sigma) {\tilde \psi} \psi - \sigma \, {\tilde \psi}^2 \psi 
  + \lambda \, {\tilde \psi} (1 + {\tilde \psi}) \psi^2 \Bigr] \ . \
\label{dpsft}
\end{equation} 
Inspection of the one-loop fluctuation corrections shows that the effective 
coupling is in fact $u = \sqrt{\sigma \lambda}$, with scaling dimension 
$[u] = \kappa^{2 - d/2}$, whence $d_c = 4$.
At least in the vicinity of $d_c$, $\lambda$ becomes irrelevant; certainly the 
ratio $\lambda / u$ scales to zero under subsequent RG transformations.
Simple rescaling ${\tilde \psi} = {\tilde \phi} \sqrt{\lambda / \sigma}$, 
$\psi = \phi \sqrt{\sigma / \lambda}$ then leaves us with the {\em effective} 
field theory\index{effective field theory}
\begin{equation}
  S_{\rm eff}[{\tilde \phi},\phi] = \int \! d^dx \, dt \, \Bigl[ 
  {\tilde \phi} \left[ \partial_t + D (r - \nabla^2) \right] \phi 
  + u \, ({\tilde \phi} \, \phi^2 - {\tilde \phi}^2 \, \phi) \Bigr] \ , 
\label{rftac}
\end{equation}
where $r = (\mu - \sigma) / D$.
This should capture the critical behavior for the non-equilibrium phase 
transition at $r = 0$ in our population dynamics model.

\begin{figure}[t]
\includegraphics[width=.42\textwidth]{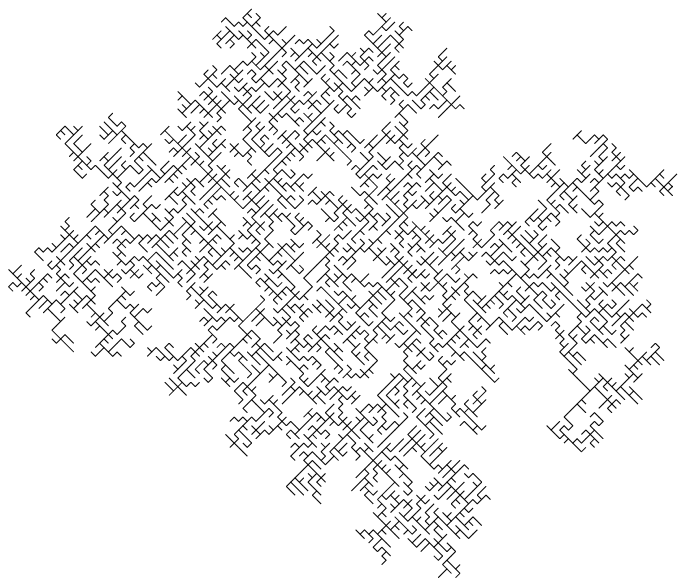}
\includegraphics[width=.16\textwidth]{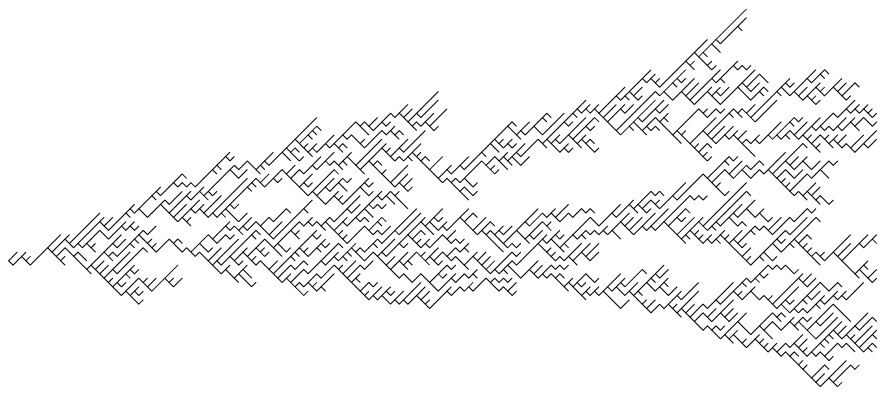}
\caption[]{Critical isotropic (a) and directed (b) percolation 
  clusters (from Ref.~\cite{frey94})}
\label{cdpcl}
\end{figure}

The action (\ref{rftac}) is known in particle physics as {\em `Reggeon' field 
theory}\index{Reggeon field theory} \cite{migdal74}.
It is invariant with respect to `rapidity inversion' 
$\phi(\vec{x},t) \to - {\tilde \phi}(\vec{x},-t)$, 
${\tilde \phi}(\vec{x},t) \to - \phi(\vec{x},-t)$.
Quite remarkably, the very same action is obtained for the threshold pair 
correlation function \cite{cardy80} in the geometric problem of {\em directed 
percolation}\index{directed percolation} (DP) \cite{kinzel83}.
Fig.~\ref{cdpcl}(b) depicts a critical directed percolation cluster, contrasted
with the structure emerging at the threshold of ordinary isotropic percolation.
At $d_c = 4$, the critical exponents\index{critical exponents} as predicted by 
mean-field theory acquire logarithmic corrections\index{logarithmic 
corrections}, and are shifted to different values by the infrared-singular 
fluctuations in $d < 4$ dimensions.
By means of the standard perturbational loop expansion in terms of 
the diffusion propagator and the vertices $\propto u$, and the 
application of the RG, the critical exponents can be computed 
systematically and in a controlled manner in a dimensional 
expansion with respect to $\epsilon = 4 - d$.
The one-loop results, to first order in $\epsilon$, as well as 
reliable values from Monte Carlo simulations in one and two 
dimensions \cite{hinrichsen00} are listed in Table~\ref{dpexp}.
Moreover, as a consequence of rapidity invariance there are only 
three independent scaling exponents, namely the anomalous field 
dimension $\eta$, the correlation length exponent $\nu$, and the 
dynamic critical exponent $z$.
All other exponents are then fixed by scaling relations, such as 
\begin{equation}
  \beta = \frac{\nu}{2} \, (z + d - 2 + \eta) = z \, \nu \, \alpha
\label{scalr}
\end{equation}
for the order parameter and critical density decay exponents.

\begin{table}
\caption{Critical exponents for the saturation density $n_s$, correlation 
  length $\xi$, time scale $t_c$, and critical density decay $n_c(t)$ for the 
  DP universality class}
\renewcommand{\arraystretch}{1.4}
\setlength\tabcolsep{5pt}
\begin{tabular}{llll}
\hline
\noalign{\smallskip}
DP exponents & $d = 1$ & $d = 2$ & $d = 4 - \epsilon$ \\
\noalign{\smallskip}
\hline
\noalign{\smallskip}
$n_s \sim |r|^\beta$ & $\beta \approx 0.2765$  & $\beta \approx 0.584$ & 
$\beta = 1 - \epsilon/6 + O(\epsilon^2)$ \\
$\xi \sim |r|^{-\nu}$ & $\nu \approx 1.100$ & $\nu \approx 0.735$ & 
$\nu = 1/2 + \epsilon/16 + O(\epsilon^2)$ \\
$t_c \sim \xi^z \sim |r|^{-z \nu}$ & $z \approx 1.576$ & $z \approx 1.73$ & 
$z = 2 - \epsilon/12 + O(\epsilon^2)$ \\
$n_c(t) \sim t^{-\alpha}$ & $\alpha \approx 0.160$ & $\alpha \approx 0.46$ & 
$\alpha = 1 - \epsilon/4 + O(\epsilon^2)$\\
\hline
\end{tabular}
\label{dpexp}
\end{table}

It is worthwhile noting that Reggeon field theory can be viewed as a dynamic 
response functional\index{reponse functional} \cite{janssen76}, and therefore 
is equivalent to an effective Langevin equation.
To this end, we integrate out the field ${\tilde \phi}$, which yields the 
statistical weight $\exp (- G)$ with 
\begin{equation}
  G[\phi] = \int \! d^dx \, dt \, \Bigl[ \partial_t \phi + D (r - \nabla^2) 
  \phi + u \, \phi^2 \Bigr]^2 \Big/ 4 \, u \, \phi \ .
\label{onmlf}
\end{equation}
After rescaling, we may interpret (\ref{onmlf}) as the Onsager-Machlup 
functional\index{Onsager-Machlup functional} associated with the Gaussian noise
distribution for the stochastic process
\begin{subeqnarray} 
  &&\partial_t \, n(\vec{x},t) = D (\nabla^2 - r) \, n(\vec{x},t) 
  - \lambda \, n(\vec{x},t)^2 + \zeta(\vec{x},t) \ ,
\label{dpleq} \\
  &&\langle \zeta(\vec{x},t) \rangle = 0 \, , \ 
  \langle \zeta(\vec{x},t) \, \zeta(\vec{x}',t') \rangle 
  = 2 \, \sigma \, n(\vec{x},t) \, \delta(\vec{x}-\vec{x}') \, \delta(t-t') \ .
\label{dplen}
\end{subeqnarray}
Here, the absorbing nature of the inactive state is reflected in the fact that 
the noise correlator is proportional to $n(\vec{x},t)$.
Of course, Eq.~(\ref{dplen}) really means that the local density is to be 
factored in when the noise average is taken, c.f.~Eq.~(\ref{onmlf}).
Alternatively, One may define 
$\zeta(\vec{x},t) = \sqrt{n(\vec{x},t)} \, \eta(\vec{x},t)$, whereupon 
$\langle \eta(\vec{x},t) \, \eta(\vec{x}',t') \rangle = 2 \, \sigma \, 
\delta(\vec{x}-\vec{x}') \, \delta(t-t')$, at the cost of introducing 
`square-root' multiplicative noise\index{multiplicative noise} into the 
Langevin equation (\ref{dpleq}).
Within the Langevin framework, we can readily generalize to arbitrary reaction 
and noise functionals $r[n]$ and $c[n]$:
\begin{subeqnarray}
  &&\partial_t \, n(\vec{x},t) = D \, \nabla^2 \, n(\vec{x},t) - 
  r[n(\vec{x},t)] + \zeta(\vec{x},t) \ ,
\label{recfn} \\
  &&\langle \zeta(\vec{x},t) \rangle = 0 \, , \
  \langle \zeta(\vec{x},t) \, \zeta(\vec{x}',t') \rangle = 
  c[n(\vec{x},t)] \, \delta(\vec{x}-\vec{x}') \, \delta(t-t') \ . 
\label{noifn}
\end{subeqnarray}
In the spirit of Landau theory, we may then expand the functionals $r[n]$ and 
$c[n]$ near the inactive phase ($n \ll 1$).
In the absence of spontaneous particle production, both must vanish at $n=0$, 
which is the condition for an absorbing state. 
Keeping only the lowest-order, relevant terms in the expansions with respect to
$n$, we thereby infer that any active to absorbing state phase transition in a 
single-species system should {\em generically} be described by 
Eqs.~(\ref{dpleq}) and (\ref{dplen}), i.e., Reggeon field theory (\ref{rftac}).
Consequently, in the absence of any special symmetries, memory effects, and
quenched disorder, we expect to find the critical exponents of the DP 
universality class \cite{grassberger78}.

\section{Variants of directed percolation processes}
\label{sec:dpva}

\begin{figure}[t]
\includegraphics[width=.40\textwidth]{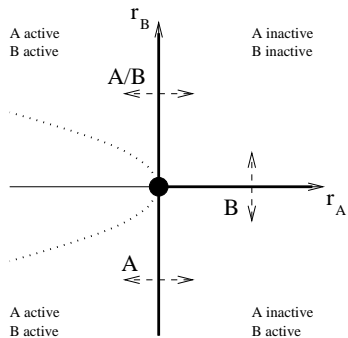}
\caption[]{Mean-field phase diagram for the two-stage unidirectionally coupled 
  DP process. The arrows indicate active to absorbing transitions for the $A$ 
  and $B$ species. The dotted parabola marks the boundary of the multi-critical
  regime \cite{tauber98}}
\label{cdppd}
\end{figure}

In fact, DP-type processes with even arbitrarily many particle species have 
been fully classified.
Consider the reactions $A \rightleftharpoons A + A$, $A \to \emptyset$, 
$B \rightleftharpoons B + B$, $B \to \emptyset$, etc., supplemented with 
bilinear couplings $A \to B + B$, $A + A \to B, \ldots$
Higher-order reactions then turn out to be irrelevant in the RG sense, and 
remarkably the critical behavior of the multi-species DP system is once again 
governed by the DP fixed point \cite{janssen01}.
However, if we allow for unidirectional {\em linear} couplings through particle
transmutations $A \to B$, $B \to C, \ldots$, with rates 
$\mu_{AB}, \mu_{BC}, \ldots$, {\em multi-critical} behavior may ensue 
\cite{tauber98}.
This becomes already manifest on the level of the coupled rate equations 
\begin{subeqnarray}
  &&\partial_t \, n_A = D \left( \nabla^2 - r_A \right) n_A 
  - \lambda_A \, n_A^2 \ , 
\label{msdpa} \\
  &&\partial_t \, n_B = D \left( \nabla^2 - r_B \right) n_B 
  - \lambda_B \, n_B^2 + \mu_{AB} \, n_A \ ,
\label{msdpb}
\end{subeqnarray}
etc.
For as long as the $A$ species is in the active phase ($r_A < 0$), the $B$ 
particle density will be non-zero as well.
As depicted in Fig.~\ref{cdppd}, this effectively `folds' half of the decoupled
$B$ transition line ($r_A < 0$, $r_B = 0$) over onto $r_A = 0$, $r_B > 0$. 
Along this half-line, the $B$ particles become `enslaved' by the $A$ species; 
and so forth further down the hierarchy of particle species.
As $r_A \to 0$ and $r_B \to 0$ simultaneously, one encounters a non-equilibrium
multi-critical point\index{multi-critical point}.
While the DP exponents $\nu$ and $z$ governing the correlation length and 
critical slowing down remain unchanged, one finds successively reduced values 
for the order parameter exponents $\beta^{(j)}$ on the $j$th hierarchy level 
\cite{tauber98}, as listed in Table~\ref{mcdpe}.
The crossover exponent associated with the multi-critical point is $\phi = 1$ 
to all orders in the $\epsilon$ expansion \cite{janssen01}.

\begin{table}
\caption{Simulation and one-loop RG results for the saturation density critical
  exponents on the first three hierarchy levels in unidirectionally coupled DP
  processes}
\renewcommand{\arraystretch}{1.4}
\setlength\tabcolsep{5pt}
\begin{tabular}{lllll}
\hline
\noalign{\smallskip}
& $d = 1$ & $d = 2$ & $d = 3$ & $d = 4 - \epsilon$ \\
\noalign{\smallskip}
\hline
\noalign{\smallskip}
$\beta^{(1)}$ & $0.280(5)$ & $0.57(2)$ & $0.80(4)$ & 
$1 - \epsilon/6 + O(\epsilon^2)$ \\
$\beta^{(2)}$ & $0.132(15)$ & $0.32(3)$ & $0.40(3)$ & 
$1/2 - \epsilon/8 + O(\epsilon^2)$ \\
$\beta^{(3)}$ & $0.045(10)$ & $0.15(3)$ & $0.17(2)$ & $1/4 - O(\epsilon)$ \\
\hline
\end{tabular}
\label{mcdpe}
\end{table}

Another mechanism to induce a different universality class in a two-species 
system is to link diffusing agents $A$ with a passive, spatially fixed, and 
initially homogeneously distributed species $X$ through the reactions 
$X + A \to A + A$ and $A \to \emptyset$.
One may then integrate out the $X$ fluctuations; upon expanding about the 
mean-field solution, the resulting effective action becomes \cite{janssen85}
\begin{eqnarray}
  &&S_{\rm eff}[{\tilde \phi},\phi] = \int \! d^dx \, dt \, \Bigl[ 
  {\tilde \phi}(\vec{x},t) \left[ \partial_t + D (r - \nabla^2) \right] 
  \phi(\vec{x},t) \nonumber \\
  &&\qquad\qquad + 2 D \, u \, {\tilde \phi}(\vec{x},t) \, \phi(\vec{x},t) 
  \int^t \!\! \phi(\vec{x},t') \, dt' - u \, {\tilde \phi}(\vec{x},t)^2 \, 
  \phi(\vec{x},t) \Bigr] \ .
\label{adynp}
\end{eqnarray}
The elimination of the passive particles $X$ thus induces memory of all 
preceding times in the particle annihilation vertex.
The DP rapidity inversion invariance is replaced by its non-local counterpart 
$\phi(\vec{x},t) \to - D^{-1} \, \partial_t {\tilde \phi}(\vec{x},-t)$, 
${\tilde \phi}(\vec{x},t) \to - D \int^{-t} \! \phi(\vec{x},t') \, dt'$.
The upper critical dimension for this {\em dynamic percolation}\index{dynamic 
percolation} universality class is shifted to $d_c = 6$.
Its static critical exponents are precisely the ones that characterize a 
critical {\em isotropic percolation}\index{isotropic percolation} cluster 
\cite{janssen85}, compare Fig.~\ref{cdpcl}(a).
To first order in $\epsilon = 6-d$, one finds $\eta = - \epsilon/21$, 
$\nu = 1/2 + 5 \epsilon/84$, $z = 2 - \epsilon/6$, and 
$\beta = \nu (d - 2 + \eta) / 2 = 1 - \epsilon/7$.
Multi-species generalizations proceed in the same way as for DP, with similar
results: 
Whereas non-linear couplings to other particle species preserve the dynamic 
universality class, a multi-critical point emerges for unidirectional particle 
transmutations, with crossover exponent $\phi = 1$ \cite{janssen01}.

\section{Diffusion-limited annihilation processes}
\label{sec:dlan}

As a preparation for the following Sec.~\ref{sec:barw}, let us now investigate 
the $k$th order {\em annihilation} reaction\index{annihilation reaction} 
$k A \to \emptyset$.
The associated rate equation reads $\partial_t \, n(t) = - \lambda \, n(t)^k$.
For radioactive decay ($k = 1$), it is naturally solved by the familiar 
exponential $n(t) = n_0 \, e^{-\lambda t}$, whereas one obtains power laws for 
$k \geq 2$, namely 
\begin{equation}
  n(t) = \left[ n_0^{1-k} + (k-1) \, \lambda \, t \right]^{-1 / (k-1)} \ .
\label{mfann}
\end{equation}
In order to consistently include fluctuations in the latter case, we again 
start out from the master equation, wherefrom we derive the action \cite{lee94}
\begin{equation}
  S[{\hat \psi},\psi] = \int \! d^dx \, dt \, \Bigl[ {\hat \psi} \, (\partial_t
  - D \, \nabla^2) \, \psi - \lambda \, (1 - {\hat \psi}^k) \, \psi^k \Bigr]\ .
\label{anmft}
\end{equation}
After performing the shift 
${\hat \psi}(\vec{x},t) = 1 + {\tilde \psi}(\vec{x},t)$, it becomes evident 
that this field theory does {\em  not} have a simple Langevin representation.
For in order to interpret ${\tilde \psi}$ as the corresponding noise auxiliary 
field, it should appear quadratically in the action only, and with negative 
prefactor.
Thus even for the pair annihilation process, the Langevin equation derived from
the action (\ref{anmft}) would entail unphysical `imaginary' noise with 
$c[n] = - 2 \, \lambda \, n^2$.

The scaling dimension of the annihilation vertex is 
$[\lambda_k] = \kappa^{2 - (k-1) d}$, whence we infer the upper critical 
dimension $d_c(k) = 2 / (k-1)$.
This leaves the possibility of non-trivial scaling behavior in low physical 
dimensions only for the pair ($k = 2$) and triplet ($k = 3$) processes.
Analyzing the field theory (\ref{anmft}) further, we see that the diffusion 
propagator does not become renormalized at all.
Consequently, $\eta = 0$ and $z = 2$ to all orders in the perturbation 
expansion.
The simple structure of the action permits summing the entire vertex 
renormalization perturbation series by means of a Bethe-Salpeter equation; in
Fourier space it reduces to a geometric series of the one-loop diagram 
\cite{lee94}.

For pair annihilation, this yields the following asymptotic behavior for the 
particle density: 
$n(t) \sim (\lambda \, t)^{-1}$, i.e., the reaction-limited power law of the 
rate equation for $d > 2$; but diffusion-limited decay $n(t) \sim (D t)^{-d/2}$
for $d < 2$.
At $d_c = 2$, one finds the logarithmic correction 
$n(t) \sim (D t)^{-1} \ln D t$.
The slower decay for $d \leq 2$ originates in the fast mutual annihilation of 
any close-by reactants; after some time has elapsed, only well-separated 
particles are left.
The annihilation dynamics thus produces {\em anti}-correlations, mimicking an 
effective repulsive interaction (which actually provides the interpretation for
the negative correlator $c[n]$).
In the ensuing {\em diffusion-limited} regime\index{diffusion-limited regime}, 
the typical particle separation scales as $\ell(t) \sim (D t)^{-1/2}$, 
whereupon indeed $n(t) \propto \ell(t)^{-d} \sim (D t)^{-d/2}$.
The same power laws hold for the pair coagulation process $2 A \to A$, albeit
with different amplitudes.
Replacing ordinary diffusion with long-range L\'evy flights\index{L\'evy 
flights} with probability $\propto r^{-d-\rho}$ of hopping a distance $r$ 
($\rho < 2$) results in $n(t) \sim (D t)^{-d / \rho}$ for $d < d_c = \rho$ 
\cite{vernon03}.
For triplet annihilation, one can similarly show that the density decays as 
$n(t) \sim (\lambda t)^{-1/2}$ for $d > 1$, with mere logarithmic corrections 
$n(t) \sim \left[ (D t)^{-1} \ln D t \right]^{1/2}$ at $d_c = 1$ \cite{lee94}.

Generalizations of the pair annihilation reaction to multiple particle types 
introduce interesting new physics.
For the {\em two-species} case $A + B \to \emptyset$ (with no concurrent 
reactions of identical particles), the rate equations read 
$\partial_t n_{A/B} = - \lambda \, n_A \, n_B$.
With equal initial densities $n_{A 0} = n_{B 0}$ they are again solved by 
$n_{A/B}(t) \sim (\lambda t)^{-1}$; however, with $n_{A 0} > n_{B 0}$, say, one
obtains $n_B(t) \sim \exp[- (n_{A 0} - n_{B 0}) \, \lambda \, t]$ for the
minority species, while the majority density saturates at $n_{A s} > 0$.
In order to establish the effects of spatial fluctuations, it is crucial to 
notice that the density difference $n_A - n_B$ remains strictly {\em conserved}
under the reactions; for $D_A = D_B$ it simply obeys the diffusion equation 
\cite{toussaint83}.
Consequently, regions with $A$ or $B$ particle excess become amplified in time.
As a result, when $n_{A 0} = n_{B 0}$, one finds that for dimensions $d \leq 4$
species segregation\index{species segregation} into $A / B$ rich domains occurs
\cite{lee95}.
The annihilation processes are then confined to sharp reaction 
fronts\index{reaction fronts}, leading to a decelerated density decay 
$n_{A/B}(t) \sim (D t)^{-d/4}$.
For unbalanced initial conditions, stretched exponential relaxation
ensues for $d < 2$: $\ln n_B(t) \sim - t^{d/2}$, whereas 
$\ln n_B(t) \sim - t / \ln t$ at $d_c = 2$ \cite{kang84}.
In one dimension, special initial configurations may change this picture:
Consider, e.g., the alternating arrangement $\ldots ABABAB \ldots$ of particles
that upon encounter react with probability one.
Now there is no reason anymore to distinguish between $A$ and $B$, and the 
system is in the $2 A \to \emptyset$ universality class.

An obvious question is then what happens for diffusion-limited pair 
annihilation of $q > 2$ particle species, with equal initial densities as well 
as reaction and diffusion rates \cite{benavraham86}.
In contrast with the two-species case, there exists no local conservation law. 
Furthermore the renormalization of the reaction vertex proceeds exactly as for 
$2 A \to \emptyset$.
Consequently, at least for $d \geq 2$, where the initial state is not that 
crucial, the long-time limit should in fact {\em generically} be governed by 
the single-species pair annihilation universality class \cite{deloubriere02}.
This is obvious for $q \to \infty$:
In this limit, the probability of like particles ever meeting vanishes, which
renders the distinction of different species meaningless.
However, in one dimension, at least in the limit of large reaction rates (which
should describe the asymptotic regime), particles of different types cannot 
pass each other. 
This topological constraint allows for species segregation to occur.
Indeed, a simplified deterministic version of the $q$-species pair annihilation
process yields \cite{deloubriere02} 
\begin{equation}
  n(t) \sim t^{- \alpha(q)} \, , \ {\rm with} \
  \alpha(q) = (q-1) / 2 q \ ,
\label{qspan}
\end{equation}
which correctly reproduces $\alpha(2) = 1/4$ and 
$\alpha(\infty) = 1/2$ in $d = 1$.
The asymptotic decay (\ref{qspan}) along with the subleading correction 
$\sim t^{-1/2}$ of the pair annihilation process without segregation were 
recently confirmed in extensive simulations \cite{zhong03}.
Yet again, special initial conditions such as $\ldots ABCDABCD \ldots$ may 
prevent segregation and instead lead to the $2 A \to \emptyset$ decay law.

\section{Branching and annihilating random walks (BARW)}
\label{sec:barw}

In order to allow again for a genuine phase transition, we combine the 
annihilation $k A \to \emptyset$ ($k \geq 2$) with branching processes 
$A \to (m+1) \, A$.
The associated rate equation for these {\em branching and annihilating random 
walks}\index{branching and annihilating random walks} (BARW) reads 
$\partial_t \, n(t) = - \lambda \, n(t)^k + \sigma \, n(t)$, with the solution 
\begin{equation}
  n(t) = n_s \Big/ \Bigl( 1 + \Bigl[ (n_s / n_0)^{k-1} - 1 \Bigr] \, 
  e^{- (k-1) \, \sigma \, t} \Bigr)^{1/(k-1)} \ .
\label{bawmf}
\end{equation}
Mean-field theory thus predicts the density to approach the saturation value 
$n_s = (\sigma / \lambda)^{1/(k-1)}$ as $t \to \infty$ for any positive 
branching rate $\sigma$.
Above the critical dimension $d_c(k) = 2/(k-1)$ therefore, the system only has 
an {\em active} phase; $\sigma_c = 0$ represents a degenerate `critical' point,
with scaling exponents essentially determined by the pure annihilation model: 
$\alpha = 1/(k-1) = \beta$, $\nu = 1/2$, and $z = 2$.
However, Monte Carlo simulations revealed a much richer picture, in low 
dimensions clearly distinguishing between the cases of {\em odd} and {\em even}
number of offspring $m$ \cite{grassberger84,takayasu92}:
For $k = 2$, $d \leq 2$, and $m$ {\em odd}, a transition to an inactive, 
absorbing phase is found, characterized by the DP critical exponents.
On the other hand, for {\em even} offspring number there emerges a phase 
transition in one dimension, described by a novel universality class with 
$\alpha \approx 0.27$, $\beta \approx 0.92$, $\nu \approx 1.6$, and 
$z \approx 1.75$.

The above mapping to a stochastic field theory, combined with RG methods, 
elucidates the physics behind those remarkable findings \cite{cardy96}.
The action for the most interesting pair annihilation case becomes 
\begin{equation}
  S = \int \! d^dx \, dt \, \Bigl[ {\hat \psi} \, (\partial_t - D \, \nabla^2) 
  \, \psi - \lambda \, (1 - {\hat \psi}^2) \, \psi^2 + \sigma \, 
  (1 - {\hat \psi}^m) \, {\hat \psi} \, \psi \Bigr] \ ,
\label{bawft}
\end{equation}
which in general allows no direct Langevin representation.
Upon combining the reactions $A \to (m+1) A$ and $2 A \to \emptyset$, one 
notices immediately that the loop diagrams generate the lower-order branching 
processes $A \to (m-1) A$, $A \to (m-3) A \ldots$
Moreover, the one-loop RG eigenvalue $y_\sigma = 2 - m(m+1)/2$ (computed at the
annihilation fixed point) shows that the reactions with smallest $m$ are the 
most relevant.
For {\em odd} $m$, we see that the generic situation is represented by $m = 1$,
i.e., $A \to 2 A$, supplemented with the spontaneous decay $A \to 0$.
After a first coarse-graining step, this latter process (with rate $\mu$) must 
be included in the effective model, which hence becomes identical with the 
action (\ref{dpmft}).
Thence we are led to Reggeon field theory (\ref{rftac}) describing the DP 
universality class, {\em provided} the induced decay processes are sufficiently
strong to render $\sigma_c > 0$.
Yet for $d > 2$ the renormalized mass term $\sigma_R - \mu_R$ remains positive,
which leaves us with merely the active phase.
For $d \leq 2$, however, the involved fluctuation integrals are 
infrared-divergent, thus indeed allowing the induced decays to overcome the 
branching processes to produce a non-trivial phase transition.
As function of dimension, the critical exponents display an unusual 
discontinuity at $d_c = 2$, as they jump from their DP to the mean-field values
as a result of the vanishing critical branching rate \cite{cardy96}.

It is now obvious why the case of {\em even} offspring number $m$ is 
fundamentally different: 
Here, the most relevant branching process is $A \to 3 A$, and spontaneous 
particle death with associated exponential decay is {\em not} generated, which
in turn precludes the previous mechanism for producing an inactive phase with 
exponential decay.
This important distinction from the odd-$m$ case can be traced to a microscopic
{\em local conservation law}\index{conservation law}, for the reactions 
$2 A \to \emptyset$ and $A \to 3 A$, $A \to 5 A \ldots$ always destroy or 
produce an even number of reactants, preserving the particle number 
{\em parity}.
Formally, this is reflected in the invariance of the action (\ref{bawft}) under
the combined inversions $\psi \to -\psi$, ${\hat \psi} \to -{\hat \psi}$.
As we saw earlier, the branching rate $\sigma$ certainly constitutes a relevant
variable near $d_c = 2$.
Therefore the phase transition can only occur at $\sigma_c = 0$, and for any 
$\sigma > 0$ there exists only an active phase, described by mean-field theory.
In two dimensions one readily computes the following logarithmic corrections: 
$\xi(\sigma) \sim \sigma^{-1/2} \ln (1/\sigma)$, and 
$n(\sigma) \sim \sigma \, [\ln (1/\sigma)]^{-2}$ \cite{cardy96}.  

\begin{figure}[t]
\includegraphics[width=.38\textwidth]{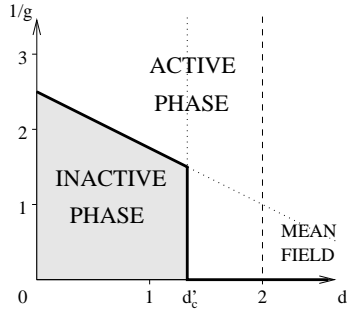}
\caption[]{Stationary states and unstable RG fixed point $1/g^*$ for BARW with 
  even offspring number (PC universality class) as function of dimension $d$ 
  \cite{cardy96}}
\label{bawphd}
\end{figure}

However, setting $m = 2$ in the one-loop value for the RG eigenvalue 
$y_\sigma$, we notice that fluctuations drive the branching vertex 
{\em irrelevant} in low dimensions $d < d_c' \approx 4/3$.
More information can be gained through a one-loop analysis at {\em fixed} 
dimension, albeit uncontrolled \cite{cardy96}.
The ensuing RG flow equations for the renormalized branching rate 
$\sigma_R = \sigma / D \kappa^2$, and annihilation rate 
$\lambda_R = C_d \lambda / D \kappa^{2-d}$, with 
$C_d = \Gamma(2-d/2) / 2^{d-1} \pi^{d/2}$ read (for $m = 2$):
\begin{equation}
  \frac{d \sigma_R}{d \ell} = \sigma_R 
  \bigg[ 2 - \frac{3 \, \lambda_R}{(1 + \sigma_R)^{2-d/2}} \biggr] , \ 
  \frac{d \lambda_R}{d \ell} = \lambda_R 
  \biggl[ 2 - d - \frac{\lambda_R}{(1 + \sigma_R)^{2-d/2}} \biggr] . \ 
\label{bawfl}
\end{equation}
The effective coupling is then identified as 
$g = \lambda_R / (1 + \sigma_R)^{2-d/2}$, which approaches the annihilation 
fixed point $g_a^* = 2 - d$ as $\sigma_R \to 0$, whereas for 
$\sigma_R \to \infty$ the flow tends towards the active state Gaussian fixed 
point $g_0^* = 0$.
The separatrix between the two phases is given by the unstable RG fixed point 
$g_c^* = 4 / (10 - 3d)$, which enters the physical regime below the borderline
dimension $d_c' \approx 4/3$, as shown in Fig.~\ref{bawphd}. 
For $d < d_c'$, this describes a dynamic phase transition with $\sigma_c > 0$. 
The aforementioned fixed-dimension RG analysis yields the rather crude values 
$\nu \approx 3 / (10 - 3 d)$, $z \approx 2$, and $\beta \approx 4 / (10 - 3 d)$
for this {\em parity-conserving} (PC) universality class.
The absence of any mean-field counterpart for this transition precludes a 
direct derivation of the `hyperscaling' relations (\ref{scalr}).
Amazingly, in this non-equilibrium system fluctuations {\em generate} 
rather than destroy an ordered phase (translating back from the domain wall to 
the spin picture) in low dimensions.
The inactive state is characterized by a vanishing branching rate, and 
consequently by the {\em algebraic} pair annihilation density decay.
For particles undergoing L\'evy flights\index{L\'evy flights}, the existence of
the power-law inactive phase is controlled by the anomalous diffusion exponent 
$\rho$, emerging for $\rho > \rho_c \approx 3/2$ in $d = 1$ \cite{vernon01}.

Invoking similar arguments for the case of triplet annihilation 
$3 A \to \emptyset$ combined with branching $A \to (m+1) A$ one would expect DP
behavior with $\sigma_c > 0$ for $m \ {\rm mod} \ 3 = 1,2$, as then the 
processes $A \to \emptyset$, $A \to 2 A$, and $2 A \to A$ are dynamically 
generated.
For $m = 3, 6, \ldots$, on the other hand, there can be different, novel 
scaling behavior, but because of $d_c = 1$ it will be limited to merely
logarithmic corrections in one dimension \cite{cardy96}.
It is also interesting to generalize the even-offspring BARW to $q$ species, 
such that only equal particles can annihilate, $A_i + A_i \to \emptyset$, but 
both reactions $A_i \to 3 A_i$ (with rate $\sigma$) and $A_i \to A_i + 2 A_j$ 
(for $j \not= i$, and with rate $\sigma'$) are possible.
It turns out that the latter process always dominates, and in fact the ratio 
$\sigma / \sigma' \to 0$ under renormalization.
Thus asymptotically one reaches the exactly analyzable $q \to \infty$ limit,
with a mere degenerate phase transition at branching rate $\sigma_c' = 0$.
Below $d_c = 2$, one finds the critical exponents $\alpha = d/2$, $\beta = 1$, 
$\nu = 1/d$, and $z = 2$ \cite{cardy96}.
The situation for $q = 1$ is thus qualitatively different from all 
multi-component cases.

\section{Annihilation--fission reactions}
\label{sec:anfr}

For single-species reactions without memory and disorder, the only remaining
processes with potentially non-mean-field scaling behavior appear to be the 
combination of purely {\em binary} annihilation and fission 
reactions\index{annihilation and fission reactions} $2 A \to A$ (with rate 
$\lambda$) and $2 A \to (m+2) A$ (rate $\sigma_m$) with $d_c = 2$, and its 
{\em triplet} counterpart ($d_c = 1$).
The former reactions subsequently generate $2 A \to (m+1) A, m A , \ldots$,
$2 (m+1) A, \ldots$, thus producing {\em infinitely} many couplings with 
identical scaling dimensions \cite{howard97}.
Upon including all these binary particle production reactions, the phase 
transition is readily seen to occur at $\lambda_c = \sum_m m \, \sigma_m$.
The inactive, absorbing phase ($\lambda > \lambda_c$) is obviously 
characterized by the power laws of the pure coagulation model.
Yet for $\lambda < \lambda_c$, the particle density diverges after a finite 
time, when no constraints on the site occupation numbers $n_i$ are imposed.
Thus the asymptotic density is finite only at the phase transition itself.
These singular features of the `bosonic' model with its highly discontinuous
phase transition are overcome by restricting the site occupation numbers to
$n_i = 0,1$.
Extensive Monte Carlo simulations have in fact revealed that this leads to a 
continuous transition, with critical exponents that seem to belong to a novel 
universality class ({\em pair contact process with diffusion}, PCPD)
\index{pair contact process with diffusion} with critical dimension $d_c = 2$
\cite{odor00}.
However, owing to the difficulty of obtaining truly asymptotic properties in
this system, where reactions become extremely rare at low densities, the 
precise nature of this critical point in purely binary reactions has remained 
elusive and rather controversial. 
This applies even to density matrix RG studies \cite{carlon01}.

It is therefore fortunate that recent work has demonstrated how to consistently
implement site occupation restrictions into the bosonic field theory 
\cite{wijland01}.
For the above binary processes, the reaction part of the action becomes
\begin{equation}
  S = \! \int \! d^dx \, dt \Bigl[ \sigma_m \, (1 - {\hat \psi}^m) \,
  {\hat \psi}^2 \, \psi^2 \, e^{- (m+2) v \, {\hat \psi} \, \psi}
  - \lambda \, (1 - {\hat \psi}) \, {\hat \psi} \, \psi^2 \, 
  e^{-2 v \, {\hat \psi} \psi} \Bigr] . \
\label{anfra}
\end{equation}
Here the exponential terms capture the occupation number limitations, with 
$[v] = \kappa^{-d}$, which suggests that $v$ represents a {\em dangerously 
irrelevant} coupling.
Indeed, consider more generally the coupled reactions $k A \to (k - l) A$ with
$0 < l \leq k$ and $n A \to (n + m) A$ with $n,m > 0$, which display a 
continuous transition for $k \leq n$.
The mean-field equations obtained from the associated actions show that site 
occupation restrictions can be neglected at low densities, yielding the 
critical exponents $\beta = 1/(k-n)$, $\nu = (k-1)/2(k-n)$, $z = 2$, and 
$\alpha = 1/(k-1)$, except for the degenerate case $k = n$, where one finds 
$m \, n_s = \ln (m \sigma_m / l \lambda)$, whence $\beta = 1$, $\nu = k/2$, 
$z = 2$, and $\alpha = 1/k$ \cite{deloubriere03,park02}.
For $k = 1$, expanding the exponentials leads to the action (\ref{dpsft}), 
which establishes that the competing processes $A \to \emptyset$, 
$A \to 2A, \ldots$ with site exclusion yield a DP phase transition.
The above field theory should also permit a systematic analysis of the 
fluctuation corrections for the purely binary and triplet reactions. 
For the latter, one expects mere logarithmic corrections at $d_c = 1$ to the 
mean-field scaling laws; yet current simulations are inconclusive 
\cite{park02}.

\section{Concluding remarks}
\label{xec:conc}

In this overview, I have outlined how non-linear stochastic processes via their
defining master equation can be represented by field theory actions, allowing 
for a thorough analysis and classification by means of the renormalization 
group.
Systems with a single `particle' species that display a non-equilibrium phase 
transition from an active to an inactive, absorbing state are generically 
captured by the directed percolation universality class.
The second prominent example, sometimes applicable when additional symmetries 
(degenerate absorbing states) are present, is the parity-conserving 
universality class of even-offspring branching and annihilating random walks.
The only other scenarios for non-trivial critical scaling behavior appear to be
provided by the solely pair or triplet annihilation--fission reactions, where 
site occupation restrictions become relevant.
The full classification of reaction-diffusion models with multiple particle 
species remains a formidable task.
Even the difference in diffusivities may become a relevant control parameter 
\cite{wijland98}.
Specifically in one dimension, exclusion constraints can play a crucial role 
\cite{odor02}.

Another obviously important open problem concerns the influence of quenched 
disorder in the reaction rates.
For example, a field theory RG investigation for DP with random percolation 
threshold yields run-away flows \cite{janssen97}, reflected in intriguing
simulation results with not entirely clear interpretation \cite{moreira96}.
A very recent strong disorder RG study, supplemented with numerical density 
matrix RG calculations, has revealed a novel disorder fixed point
\cite{hooyberghs03}.
A better understanding of spatially varying reaction rates might also explain 
the conspicuous rarity of clear-cut experimental realizations even for the 
supposedly ubiquitous DP universality class \cite{hinrichsen0b}.
In fact the single verification of DP scaling behavior appears to be its 
observation in spatio-temporal intermittency in ferrofluidic spikes 
\cite{rupp02}.
Thus many intriguing issues are still open; but I expect that in addition to 
increasingly more extensive Monte Carlo simulations and sophisticated numerical
techniques, field theory representations and subsequent analysis by means of 
the renormalization group will remain an invaluable tool for the further 
understanding of cooperative phenomena and scale-invariance in interacting 
non-equilibrium systems.

\noindent {\em Acknowledgements:} 
I am grateful for the opportunity to present this over\-view at the 2003 DPG 
Spring Meeting.
I gladly acknowledge fruitful collaboration and discussions with John Cardy, 
who introduced me to the topics discussed here, as well as with Olivier 
Deloubri\`ere, Erwin Frey, Yadin Goldschmidt, Manoj Gopalakrishnan, 
Geoff Grinstein, Malte Henkel, Henk Hilhorst, Haye Hinrichsen, Martin Howard, 
Hannes Janssen, Miguel Mu\~noz, G\'eza \'Odor, Klaus Oerding, Zolt\'an R\'acz, 
Beth Reid, Beate Schmittmann, Gunter Sch\"utz, Franz Schwabl, Steffen Trimper, 
Ben Voll\-mayr-Lee, Fr\'ed\'eric van Wijland, and Royce Zia.       
This research is presently funded through the National Science Foundation 
(grant DMR 0075725) and the Jeffress Memorial Trust (J-594).
Earlier support came from the EPSRC (GR/J78327), the European Commission 
(ERB FMBI-CT96-1189), and the DFG (Ta 177/2).

%
\clearpage
\addcontentsline{toc}{section}{Index}
\flushbottom
\printindex

\end{document}